# Optical Force on Two-level Atoms by Few-cycle Pulsed Gaussian Laser field beyond the Rotating Wave Approximation

Parvendra Kumar and Amarendra K. Sarma*
Department of Physics, Indian Institute of Technology Guwahati, Guwahati-781039, Assam, India.
*Electronic address: aksarma@iitg.ernet.in

**Abstract:** We report a study on light force on a beam of neutral two-level atoms superimposed upon a few-cycle pulsed Gaussian laser field under both resonant and off-resonant condition. The phenomena of focusing, defocusing and steering of the neutral atoms in the laser field is analysed by solving the optical Bloch equation beyond the rotating wave approximation and the force equation self-consistently .We find that two-level atoms in an atomic beam could be focused and defocused for large, positively and negatively detuned interaction even in the regime of extreme nonlinear optics. The so-called optical potential may be used for stable trapping of the neutral two-level atoms for large positively detuned interaction. This work successfully reproduces some of the features reported in recent experimental and theoretical works.
*PACS number(s)*: 37.10.Vz, 42.50.Hz, 33.15.−e, 42.50.Tx

## I.   Introduction

Since the appearance of the classic paper by Ashkin [1] on atomic beam deflection by resonance-radiation pressure force owing to laser light, a significant amount of theoretical and experimental studies on resonance-radiation force is carried out in various contexts [2-16]. Trapping of atoms by resonance radiation pressure force due to continuous-wave (cw) light, followed by experimental demonstration of focusing of neutral atoms [3] , the field of atom or molecule trapping have virtually got exploded and it still remains an interesting area of research [4-6]. It should be noted that the mechanical effects of light, particularly laser beam, on particles such as atoms, molecules, ions etc. have been successfully exploited in as diverse areas as optical tweezers [7], atom optics [8], Bose–Einstein condensation [9], laser cooling and trapping [4], quantum information [10], etc. Many authors have studied the radiation forces exerted on neutral atoms [11-16]. In this context, the two-level atomic system may be the most studied one [17]. The forces on two-level atoms are generally calculated by using the steady-state solutions of the so-called optical Bloch equations within the rotating wave approximation (RWA) [18]. However, the rotating wave approximation is questioned owing to the recent progress in the generation of intense femtosecond and attosecond optical pulses [19-26]. It is now widely accepted that when one has to deal with intense few cycle pulses, rules of the so called traditional nonlinear optics or even quantum optics have to be re-examined, modified or corrected. The extremely important slowly varying envelope approximation (SVEA) in nonlinear optics or the so called rotating wave approximation (RWA) in quantum optics can no longer be used in this new regime [20]. It is to be noted that recent experiments on semiconductors have shown that in the regime of extreme nonlinear optics, where $\Omega_R/\Omega \approx 1$ and $\Omega_R/\Omega \gg 1$, the description of atomic system in terms of two-level systems has been able to reproduce the experimental results amazingly well [25]. $\Omega_R$ is the peak of the Rabi frequency and $\Omega$ is the transition frequency between the two levels. As many new phenomena is predicted and discovered in this new regime of optics, also known as extreme nonlinear optics [25], it may be quite interesting to relook or re-examine the so-called light or optical force on a two-level atomic system in a few-cycle pulsed laser field. In fact, recently Lembessis and Ellinas [27] have carried out a theoretical analysis in the context of optical dipole trapping beyond the RWA. Their analysis is based on the Heisenberg operator perturbation techniques, rather than the optical Bloch equations. In this work we calculate the light force on a beam of neutral two-level atomic beam superimposed upon a few-cycle pulsed Gaussian laser fields. Then we carry out a numerical study of the appropriate optical Bloch equation without invoking the RWA and the force expression in a self-consistent manner. The phenomena of focusing, defocusing and steering of the neutral atoms owing to the light force due to atom-field interaction are analyzed briefly. This work is largely motivated by the work of Bjorkholm *et al.* [3] where they carried out experimental studies of focusing of sodium atoms by using a cw beam tuned near an atomic resonance. Our analysis is loosely based on the treatment by R. J. Cook [11]. It may be noted that the theory of atomic motion in resonant electromagnetic wave proposed by R. J. Cook [11-12] within the RWA successfully explains the nature of the resonant radiation forces. This theory, however, may fail in the regime of extreme nonlinear optics due to the limitations of RWA where the Rabi frequency becomes comparable to or even larger than the transition frequency of the two level atoms. In fact we have shown numerically that atoms can be steered for resonant interaction only, if the Rabi frequency is less compared to the transition frequency of the atoms. In the regime of extreme nonlinear optics, the force component in the direction of atomic motion becomes oscillatory and the atoms can no longer be steered. Our discussion of the nature of the light force, which arises due to atom-field interaction, for both resonant as well as for non-resonant interaction is based on the optical Bloch equation without invoking the RWA.

## II. Theory

Within the electric dipole approximation, the interaction between a system of two level atoms with a superimposed and co-propagating collimated classical laser field is described by the following Hamiltonian [28]:

$$H = \frac{\vec{P}^2}{2M} + H_0 - \vec{\mu}.\vec{E}(\vec{R},t). \tag{1}$$

Here $\vec{P}$ is the centre-of-mass momentum operator of the atoms, $\vec{\mu}$ is the atomic dipole moment operator and $\vec{E}(\vec{R},t)$ is the electric field of the laser field evaluated at the centre-of-mass (CM) position $\vec{R}$ of the atoms. On the right hand side of equation (1), the first term is the CM kinetic energy of the moving atoms; the second one refers to the internal energy Hamiltonian of the unperturbed atoms while the last term describes the interaction energy between the atoms and light field within the electric dipole approximation. In the so called Raman-Nath approximation [28], the approximate Hamiltonian is given by the following equation:

$$H = H_0 - \vec{\mu}.\vec{E}(\vec{R},t) \tag{2}$$

Raman Nath approximation is valid in the present study due to our consideration that the interaction energy is very large compared to the centre of mass kinetic energy and the interaction time involved is of the order of a few femtosecond. The change in momentum of the atoms is given by the following Heisenberg equation's of motion:

$$\frac{d\vec{P}}{dt} = \frac{i}{\hbar}\left[H,\vec{P}\right] = \vec{\nabla}_R\left(\vec{\mu}.\vec{E}\right) \tag{3}$$

The light force acting on the atomic centre of mass can be thought as the expectation value of equation (3). Setting $\vec{r} = \langle\vec{R}\rangle$, we obtain the average force on the atoms as follows:

$$\vec{F} = M\ddot{\vec{r}} = \left\langle \frac{d\vec{P}}{dt} \right\rangle = \left\langle \vec{\nabla}(\vec{\mu}.\vec{E}) \right\rangle. \tag{4}$$

Considering the laser field in the following form:

$$\vec{E}(\vec{r},t) = \hat{\eta} A(\vec{r},t)\cos(\phi(z) - \omega t), \tag{5}$$

where $A(\vec{r},t)$ is the envelope, $\hat{\eta}$ is the polarization direction, $\phi(z)$ is the phase and $\omega$ is the operating frequency of the laser field. Then from Eq. (4) and Eq. (5), assuming $\vec{\nabla}|\vec{E}(\vec{r},t)|$ to be uniform across the atomic wave packet, we obtain:

$$\vec{F} = \langle \vec{\mu}.\hat{\eta} \rangle \left[ \{\vec{\nabla}A(\vec{r},t)\}\cos(\phi(z) - \omega t) - \vec{\nabla}\phi(z)A(\vec{r},t)\sin(\phi(z) - \omega t) \right] \tag{6}$$

The expectation value in equation (6) can be written in terms of density matrix as: $\langle \vec{\mu}.\hat{\eta} \rangle = \mu(\rho_{12} + \rho_{21}) = \mu u$, where $u$ is the well-known Bloch vector component which accounts for the dispersive effects of the two-level atomic medium. $\rho_{12}$ and $\rho_{21}$ are the off-diagonal elements of the density matrix with $|1\rangle$ and $|2\rangle$ referring to the ground and the excited state of the two-level atom respectively. The Bloch vector component $u$ is well described by the following optical Bloch equations [25]:

$$\begin{aligned} \frac{du}{dt} &= \Omega v - \frac{u}{T_2} \\ \frac{dv}{dt} &= \Omega u - 2\Omega_R(\vec{r},t)w - \frac{v}{T_2} \\ \frac{dw}{dt} &= 2\Omega_R(\vec{r},t)v - \frac{(w+1)}{T_1} \end{aligned} \tag{7}$$

where $u, v$ and $w$ are the three components of the Bloch vector. $T_2$ and $T_1$ are respectively the dipole-dephasing and spontaneous decay time. As the atom-field interaction time, due to extremely short duration of the few-cycle laser field, is negligibly small compared to $T_1$, the terms associated with it could be neglected [23-24]. $\Omega$ is the transition frequency of the two level atoms. $\Omega_R(\vec{r},t)$ is the Rabi frequency, defined as $\Omega_R(\vec{r},t) = \vec{\mu}.\vec{E}(\vec{r},t)/\hbar$. The so-called detuning parameter, to be used later in this work, is defined as $\Delta = \Omega - \omega$. It may be noted that the optical Bloch equations are written without invoking RWA. So, in terms of Bloch vector we can write the expression for light force as follows:

$$\vec{F} = \mu u \left[ \{\vec{\nabla}A(\vec{r},t)\}\cos(\phi(z) - \omega t) - \vec{\nabla}\phi(z)A(\vec{r},t)\sin(\phi(z) - \omega t) \right]. \tag{8}$$

We find that the light force is explicitly dependent on the $u$-component of the Bloch vector unlike previous expressions for the light force, derived under the RWA, where the force was found to depend both on $u$ and $v$ component of the Bloch vector [25]. We may interpret this difference in results physically as follows: In the RWA approximation, the light force is generally expressed as the sum of two forces, namely, the reactive force and the dissipative force [18]. The reactive force, being proportional to the $u$-component of Bloch vector, does not involve absorption of energy from the laser field. Rather, it is solely due to the exchange and redistribution of momentum between the atoms and various plane waves composing the laser field. On the other hand, the dissipative force, proportional to the $v$-component of Bloch vector, is related to the absorption and emission of energy. The dissipative force arises from the impulse experienced by an atom when it absorbs or emits a quantum of photon momentum. The light force expression that we have derived does not depend on the $v$- component of Bloch vector explicitly owing to the fast laser-atom interaction compared to the slow spontaneous process and the non-RWA treatment of forces in the regime of extreme nonlinear optics. So the light force that we have got is conservative and solely due to the interaction of the two-level atoms with the gradient of the electric field envelope and of the phase. Now we consider a few-cycle pulsed Gaussian laser field propagating along the z-direction described by the following equation:

$$\vec{E}(\vec{r},t) = \hat{\eta} E_0 \exp\left[-\left\{\left(\frac{x^2+y^2}{\omega_0^2}\right)+\left(\frac{t^2}{\tau^2}\right)\right\}\right] \cos(kz-\omega t) \tag{9}$$

where $E_0$ is the peak amplitude, $\omega_0$ is the beam waist and $k$ is the wave-vector of the Gaussian laser field. $\tau$ is the temporal pulse-width, related to the full-width at half maxima (FWHM) of the laser field by $\tau_p = 1.177\tau$. From Eq. (8) and (9) we obtain the transverse and longitudinal component of the light force as follows:

$$\begin{aligned} F_x &= -\frac{2\mu E_0 \, xu}{\omega_0^2} \exp\left[-\left\{\left(\frac{x^2+y^2}{\omega_0^2}\right)+\left(\frac{t^2}{\tau^2}\right)\right\}\right] \cos(kz-\omega t) \\ F_y &= -\frac{2\mu E_0 \, yu}{\omega_0^2} \exp\left[-\left\{\left(\frac{x^2+y^2}{\omega_0^2}\right)+\left(\frac{t^2}{\tau^2}\right)\right\}\right] \cos(kz-\omega t) \\ F_z &= -\mu k E_0 u \exp\left[-\left\{\left(\frac{x^2+y^2}{\omega_0^2}\right)+\left(\frac{t^2}{\tau^2}\right)\right\}\right] \sin(kz-\omega t) \end{aligned} \tag{10}$$

The so-called optical potential [27], defined by $\vec{F} = -\vec{\nabla} U$, associated with the light force could be easily expressed by the following equation:

$$U = -\mu E_0 u \exp\left[-\left\{(x^2+y^2)/\omega_0^2 + (t/\tau)^2\right\}\right] \cos(kz-\omega t) \tag{11}$$

In order to derive the above expression for optical potential we have assumed that the spatial variation of $u$ is negligible.

### III. Results and Discussions:

In order to understand the temporal evolution of the light force on the two-level neutral atoms, using the few-cycle pulsed Gaussian laser field described by Eq. (9), we solve Eq. (7) and (10) numerically. We assume the atoms to be at the ground state initially and the beam to be focussed at $z=0$. We compare both the RWA and non-RWA cases. We find that the phenomena of focusing, defocusing and steering of the atomic beam may occur depending on the detuning parameter and the peak Rabi frequency. The peak Rabi frequency is defined as the Rabi frequency at $r=0$ and $t=0$. In the rest of the work, the Rabi frequency used refers to the peak Rabi frequency. We choose the following parameters for our numerical calculations: $\tau_p = 13.8$ fs, $\Omega = 2.2758$ rad/fs, $T_2 = 200$ fs, $\mu = 2.65\,e\,\overset{\circ}{A}$ and $\omega_0 = 1\ \mu$m. The x- and y- component of the light force is termed as the transverse force, $F_T$, while the z-component is termed as the longitudinal force, $F_L$. Both these forces are calculated at

$x, y = 0.5\,\mu$m. Fig. 1 depicts the temporal evolution of the light force on the two-level atoms in an atomic beam at different Rabi frequencies for $\Delta = 1.7758$ rad/fs.

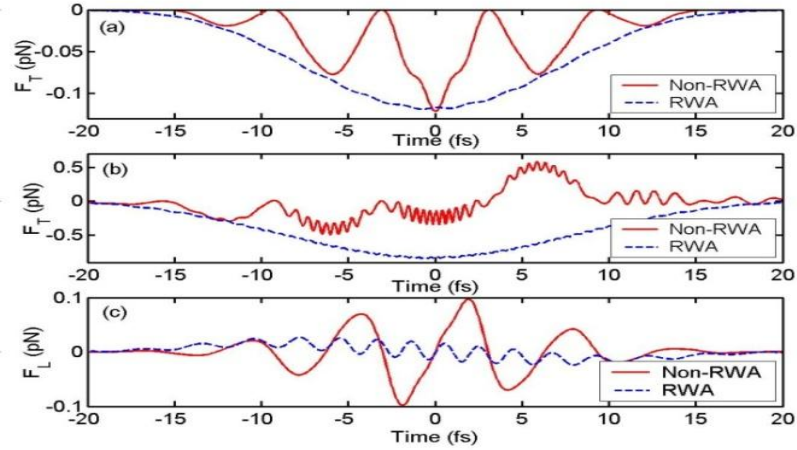

Fig. 1 (Color online): Temporal evolution of light force: (a) Transverse force vs. Time for $\Omega_R = 2.2758$ rad/fs (b) Transverse force vs. Time for $\Omega_R = 14.2758$ rad/fs and (c) Longitudinal force vs. Time for $\Omega_R = 2.2758$ rad/fs

Fig. 1 (a) shows that, the transverse force being negative, the two level atoms in the atomic beam experiences a net attractive force which may result in focusing of the atoms around $z = 0.$ Both the RWA and non-RWA treatment qualitatively predicts the same result in the limit $\Omega \approx \Omega_R$ and large detuning. However, as the Rabi frequency is increased further, for $\Omega_R > \Omega$, the non-RWA treatment of the transverse force deviates from that of RWA. As could be observed from Fig. 1(b), the non-RWA transverse force may become positive or negative with time. The time averaged longitudinal force is found to be nearly zero, as could be seen from Fig. 1(c). Next, we consider the case of negative detuning with, say $\Delta = -1.7758$ rad/fs. Fig.2 exhibits the temporal evolution of the transverse light force for different Rabi frequencies. It should be noted that in Fig. 2, in order to get proper scaling, we have reduced the RWA force by a factor of three.

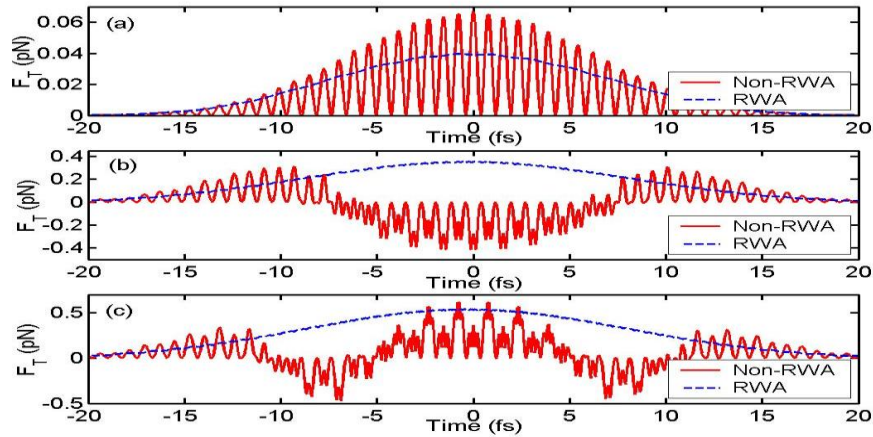

Fig. 2 (Color online): Temporal evolution of transverse light force: (a) $\Omega_R = 2.2758$ rad/fs (b) $\Omega_R = 18.2758$ rad/fs and (c) $\Omega_R = 28.2758$ rad/fs

The RWA treatment of the transverse force shows that its temporal evolution is independent of the Rabi frequency and is always positive. So the atomic beam would experience a net repulsive force leading to the phenomena of defocusing. On the other hand, the non-RWA treatment shows significant deviation in results. As could be seen from Fig. 2(b) and (c), with increase in the Rabi frequency beyond $\Omega_R > \Omega$, the transverse force exhibits both the attractive and repulsive feature. So we may conclude that time controlled focusing and defocusing of atoms may be achieved with large negative detuning. This might enable us to deposit atoms onto a substrate in a controllable way with judicious choice of Rabi frequency and the detuning parameter. The resonant case, which seems to be the most studied one in literature [18], is discussed next. Fig. 3 depicts the temporal evolution of the transverse force at different Rabi frequencies under resonant condition, i.e. $\Delta = 0$ while Fig.4 depicts the temporal evolution of the longitudinal force. It is clear from Fig. 3 that under resonant condition the transverse component of the light force vanishes for RWA. The non-RWA transverse force shows oscillatory behaviour and when taken time-average it vanishes. The fact that the transverse component of the light force vanishes within RWA under the resonant condition is supported by the analytical calculations reported in Ref. [16].

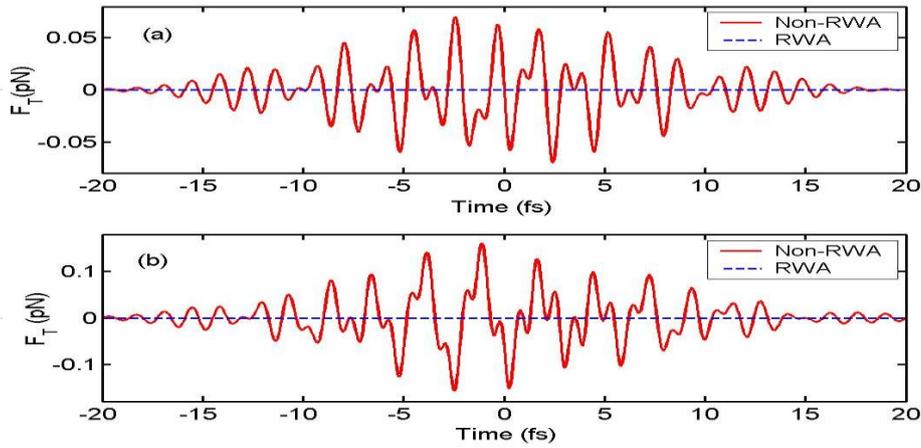

Fig. 3 (Color online): Temporal evolution of transverse light force: (a) $\Omega_R = 2.2758$ rad/fs (b) $\Omega_R = 4.2758$

Fig. 4(a) shows that the longitudinal component of the light force is non-zero when $\Omega_R < \Omega$, a regime where RWA may be valid. So steering of atoms may be possible in this regime. However, as the Rabi frequency is increased, the time-averaged longitudinal force vanishes and steering of atoms may no longer be possible for $\Omega_R \geq \Omega$. It is worthwhile to note that the peak amplitude of the electric field cannot be increased arbitrarily; thereby increasing the

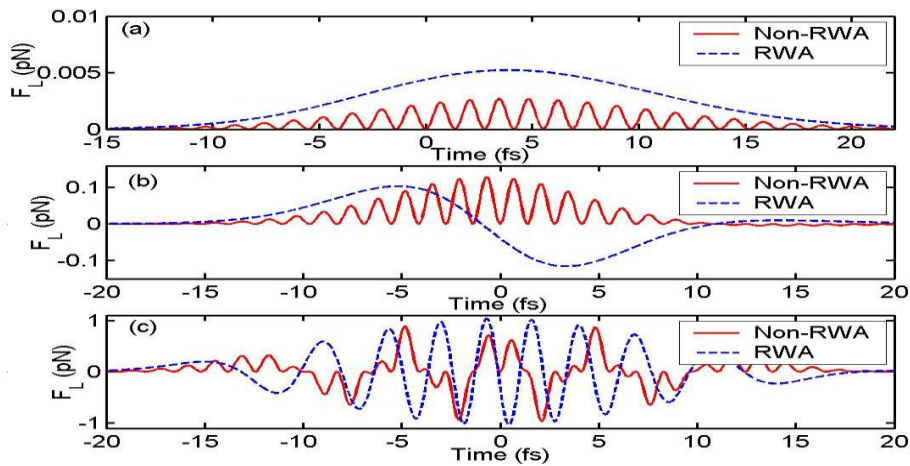

Fig. 4 (Color online): Temporal evolution of longitudinal light force: (a) $\Omega_R = 0.02758$ rad/fs (b) $\Omega_R = 0.2758$ rad/fs and (c) $\Omega_R = 2.2758$ rad/fs

Rabi frequency, because if the peak amplitude of electric field becomes comparable to the electric field strength of the atom, it may get ionized and the description of two level atoms by optical Bloch equation would no longer be valid. Finally, in Fig. 5 (a) and (b) we plot the spatio-temporal profile of the optical potential for $\Omega_R < \Omega$ and $\Omega_R = \Omega$ respectively with $\Delta = 1.7758$ rad/fs, while Fig. 5(c) depicts the case for $\Omega_R > \Omega$ with $\Delta = 1.9578$ rad/fs.

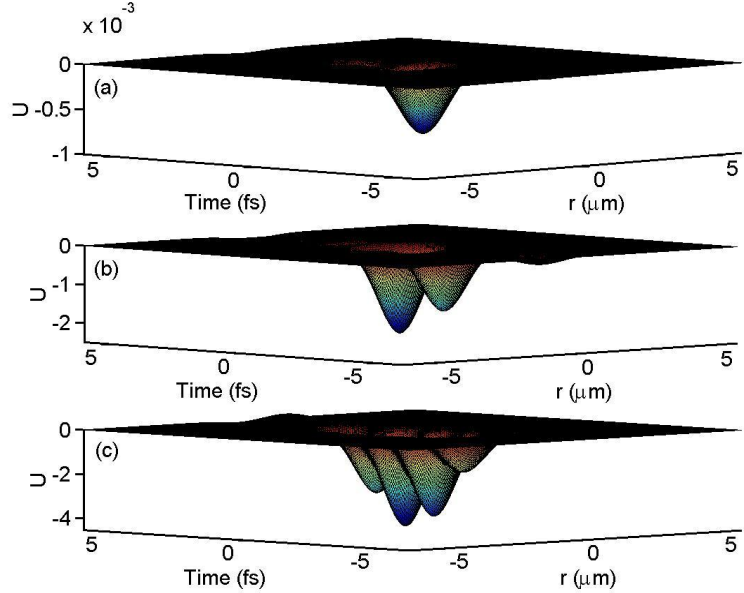

Fig. 5 (Color online): Spatio-temporal profile of the optical potential (a) $\Omega_R = 0.02758$ rad/fs (b) $\Omega_R = 2.2758$ rad/fs and (c) $\Omega_R = 4.2758$ rad/fs. Here $r = \sqrt{x^2 + y^2}$.

A careful look at Fig. 5(a) and (b) reveals that the optical potential is negative, mainly around $r, t = 0$ and its magnitude increases with increase in the Rabi frequency. So, two-level atoms in an atomic beam may be trapped by the time dependent optical potential even in the regime of extreme nonlinear optics. However, if we increase $\Omega_R$ further, keeping $\Delta$ fixed, the optical potentials oscillate between positive and negative values, a feature that could be understood from Fig.1 intuitively. We find that the optical potential could be kept negative by enhancing the $\Delta$-parameter, as indicated by Fig. 5(c). One may note that for $\Omega_R > \Omega$, as could be seen from Fig. 5(c), the optical potential becomes negative in other temporal regime as well. This may give us a tool to manipulate the optical trap in the regime of extreme nonlinear optics. Fig. 6 depicts the spatial variation of the optical potential for various Rabi frequencies with $\Delta = 1.7758$ rad/fs. It may be worthwhile to mention that the nature of the optical potential reported in this work matches quite well with that of the recent experimental work, in the context of trapping of nanoparticles with femtosecond pulses, reported in Ref. [7]. The optical potential is getting split with the increase of the Rabi frequency. One may note the absence of splitting of the optical potential in Fig. 5 as against the one in Fig. 6. This difference is occurring owing to the fact that in Fig. 5, the $u$-component of the Bloch vector is taken to be spatially independent while spatial dependency is taken into account while plotting Fig. 6. As *u* is related to the polarization of the atoms, our argument may be justified by the similar one provided in Ref. [7].

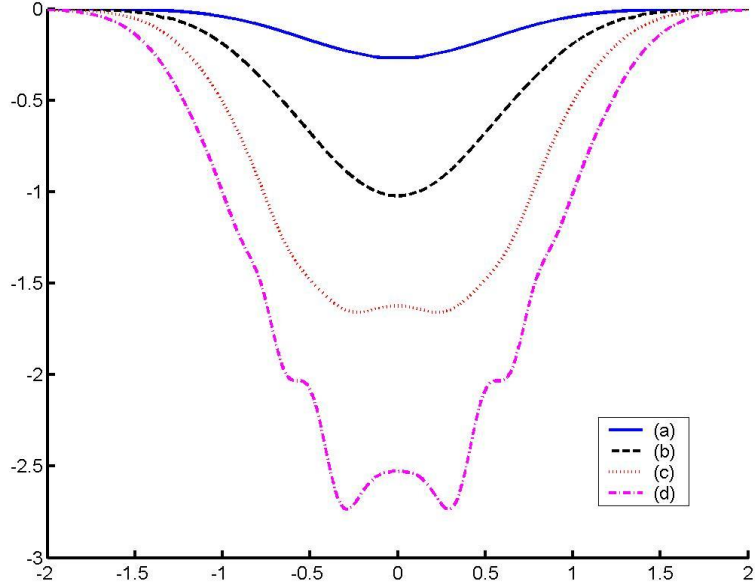

Fig. 6 (Color online): Spatial profile of the optical potential (a) $\Omega_R$=0.6379 rad/fs (b) $\Omega_R$=1.2758 rad/fs (c) $\Omega_R$=2.2758 rad/fs and (d) $\Omega_R$=4.2758 rad/fs

## IV. Conclusion

To conclude, we have studied the light force on a beam of neutral two-level atoms superimposed upon a few-cycle pulsed Gaussian laser field under both resonant and off-resonant condition. A rigorous numerical study is carried out to analyse focusing, defocusing and steering of the neutral atoms in the laser field. We find that two-level atoms in an atomic beam could be focused and defocused for large, positively and negatively detuned interaction even in the regime of extreme nonlinear optics. The so-called optical potential may be used for stable trapping of the neutral two-level atoms for large positively detuned interaction. We find that the light force beyond the RWA has turned out to be conservative one for the particular problem considered in this work and so it cannot be used to cool a sample of two-level atoms. The treatment based on Ehrenfest's theorem [11] describes only the mean light force and is silent about the fluctuations of the light force about its mean value. This, however should not reduce the effectiveness of our work as our main objective in this work was to get a comprehensive idea about the light force on an atomic beam of two-level atoms beyond RWA superimposed upon a few-cycle pulsed Gaussian laser field. Moreover this work successfully reproduces some of the features reported in recent experimental and theoretical works [7, 27]. This work could be extended to three-level atomic systems as well, where one may explore the possibility of manipulating the whole trapping mechanism using the so-called control field.